%Paper: hep-th/9410014
%From: Martin Cederwall <tfemc@fy.chalmers.se>
%Date: Tue, 4 Oct 94 10:44:35 +0100
%Date (revised): Tue, 4 Oct 94 12:32:02 +0100

%%%%%%%%%%%%%%%%%%%%%%%%%%%%%%%%%%%%%%%%%%%%%%%%%%%%%%%%%%%%%%%%%
%								%
%	Define script letters as rsfs		 		%
%		(or redefine as cal)				%
%		 						%
% 								%
%%%%%%%%%%%%%%%%%%%%%%%%%%%%%%%%%%%%%%%%%%%%%%%%%%%%%%%%%%%%%%%%%
\newfam\scrfam
\batchmode\font\tenscr=rsfs10 \errorstopmode
\ifx\tenscr\nullfont
	\message{rsfs script font not available. Replacing with calligraphic.}
\else	\font\sevenscr=rsfs7
	\font\fivescr=rsfs5
	\skewchar\tenscr='177 \skewchar\sevenscr='177 \skewchar\fivescr='177
	\textfont\scrfam=\tenscr \scriptfont\scrfam=\sevenscr
	\scriptscriptfont\scrfam=\fivescr

\fi
%%%%%%%%%%%%%%%%%%%%%%%%%%%%%%%%%%%%%%%%%%%%%%%%%%%%%%%%%%%%%%%%%
%								%
%	Euler fraktur (or redefine as italic)			%
%								%
%%%%%%%%%%%%%%%%%%%%%%%%%%%%%%%%%%%%%%%%%%%%%%%%%%%%%%%%%%%%%%%%%
\newfam\frakfam
\batchmode\font\twelvefrak=eufm10 scaled\magstep1 \errorstopmode
\ifx\twelvefrak\nullfont
	\message{Euler fraktur not available. Replacing with italic.}
\else	\font\tenfrak=eufm10 \font\sevenfrak=eufm7 \font\fivefrak=eufm5
	\textfont\frakfam=\tenfrak
	\scriptfont\frakfam=\sevenfrak \scriptscriptfont\frakfam=\fivefrak
	
\fi
%%%%%%%%%%%%%%%%%%%%%%%%%%%%%%%%%%%%%%%%%%%%%%%%%%%%%%%%%%%%%%%%%
%								%
%	Blackboard bold (or redefine as boldface)		%
%								%
%%%%%%%%%%%%%%%%%%%%%%%%%%%%%%%%%%%%%%%%%%%%%%%%%%%%%%%%%%%%%%%%%
\newfam\msbfam
\batchmode\font\twelvemsb=msbm10 scaled\magstep1 \errorstopmode
\ifx\twelvemsb\nullfont\def\Bbb{\bf}
	\message{Blackboard bold not available. Replacing with boldface.}
\else	\catcode`\@=11
	\font\tenmsb=msbm10 \font\sevenmsb=msbm7 \font\fivemsb=msbm5
	\textfont\msbfam=\tenmsb
	\scriptfont\msbfam=\sevenmsb \scriptscriptfont\msbfam=\fivemsb
	\def\Bbb{\relax\ifmmode\expandafter\Bbb@\else
 		\expandafter\nonmatherr@\expandafter\Bbb\fi}
	\def\Bbb@#1{{\Bbb@@{#1}}}
	\def\Bbb@@#1{\fam\msbfam\relax#1}
	\catcode`\@=\active
\fi
\font\eightcp=cmcsc8

\font\twelvebf=cmbx10 scaled\magstep 1
\font\sevenit=cmti7
\hsize=15cm
\hoffset=1cm
\vsize=21cm
\voffset=.5cm

\baselineskip=15pt
\vsize=21cm
\hsize=15cm
\voffset=1cm
\hoffset=1cm

\nopagenumbers

\headline={\ifnum\pageno=1\hfill\else\hfill
{\eightcp Why Don't We Have a Covariant Superstring Field Theory?}\hfill\fi}
\def\makeheadline{\vbox to 0pt{\vss\noindent\the\headline\break
\hbox to\hsize{\hfill}}
	\vskip\baselineskip}

\def\vs{\vskip\parskip}

\def\!{\mskip-.7\thinmuskip}

\def\R{\Bbb R}
\def\C{\Bbb C}
\def\H{\Bbb H}
\def\O{\Bbb O}
\def\K{\Bbb K}

\def\pro#1{\!\Buildrel
	\raise 4pt\hbox{\the\scriptscriptfont0 #1}\under\circ\!}

\def\first{$1^{\scriptscriptstyle\underline{\underline{st}}}$}
\def\second{$2^{\scriptscriptstyle\underline{\underline{nd}}}$}
\def\brk{\hfill\break}
\def\half{{1\over 2}}

\noindent{\twelvebf Why Don't We Have a}\brk
\noindent{\twelvebf Covariant Superstring Field Theory?}
\vskip-6\baselineskip
\hbox to\hsize{\hfill G\"oteborg-ITP-94-24}
\hbox to\hsize{\hfill{\tt hep-th/9410014}}
\vskip7\baselineskip
\noindent Martin Cederwall
\vskip.4cm
\noindent Institute for Theoretical Physics\brk
\noindent Chalmers University of Technology and G\"oteborg University
	\brk
\noindent S-412 96 G\"oteborg, Sweden
\vs\catcode`\@=12
\noindent email: {\it tfemc@fy.chalmers.se}
\catcode`\@=11
\vskip.7cm
\centerline{\sevenit Talk presented at the 1st Feza G\"ursey Memorial
	Conference on Strings and Symmetries, Istanbul, June 1994,}
\vskip-6pt
\centerline{\sevenit and at the 28th
	International Symposium on the Theory of Elemaentary Particles,
	Wendisch-Rietz, September 1994.}
\vskip.7cm
\noindent{\bf Abstract. }This talk deals with the old problem of formulating
a covariant quantum theory of superstrings, ``covariant'' here meaning
having manifest Lorentz symmetry and supersymmetry.
The advantages and disadvantages of several quantization methods
are reviewed. Special emphasis
is put on the approaches using twistorial variables, and the algebraic
structures of these.
Some unsolved
problems are identified.
\vskip1cm

Before going into supersymmetric strings, let me examine the situation
in bosonic string theory.

The bosonic string has a well defined (Lorentz) covariant \first\
quantized formulation as a gauge theory, preferably through BRST [1].
This formulation is the starting point and an absolute
prerequisite for the \second\ quantization, the {\it field theory} [2].
It is probable that field theory can provide a framework for posing questions
about the big symmetries of string theory, including general
coordinate invariance. Some aspects of background invariance have
already been addressed in bosonic string theory [3].

In this perspective, what is the corresponding status of superstring theory?
It is not so good. Why is this so?

To be clear about the ambitions, one would like a covariant quantum superstring
theory to fulfill the following requirements:
\item{1.}{It should have manifest space-time symmetry, including
	supersymmetry.}
\item{2.}{It should contain \first\ class constraints only.}

\noindent Concerning the second of these points,
there may well be \second\ class
constraints present, but they have to be dealt with in a covariant manner
before \first\ quantization. There are different methods for doing this --
in Dirac's original treatment of constraints [4] \second\ class constraints
are eliminated consistently by letting remaining variables (parametrizing
the \second\ class constraint surface) obey Dirac brackets; in the method
by Batalin and Fradkin [5] additional constraints are added to turn
the constraints into \first\ class ones.

There are a number of quite different formulations of superstring theory.
Let us examine them with respect to the requirements! The three main
classes of models are
\item{1.}Spinning string
\item{2.}Green-Schwarz superstring
\item{3.}Twistor superstrings (main subject of this talk).

\noindent In each of these approaches, there may be alternative formulations
or modifications. I will briefly review the different models, and
make some indications on to what extent and on which points
the requirements we have
set up fail to be fulfilled.

The spinning/fermionic/NSR
string [6] has an $N\!=\!1$ world-sheet supersymmetry.
The action is the $N\!=\!1$ generalization of
$$
S=\int d^2\sigma\sqrt{-g}g^{\alpha\beta}
	\partial_\alpha X^\mu\partial_\beta X_\mu
$$
where a world-sheet gravitino and a fermionic space-time vector
have been included [7].
The {\it space-time} supersymmetry of the spinning string is
not present until the GSO projection [8] is performed on the spectrum
and it is highly non-manifest.
A functioning field theory exists for the spinning string [9], but it
has of course the same drawbacks as the \first\ quantized theory.
The only thing that this formulation does not give us is manifest
supersymmetry. The calculational power that two-dimensional conformal
field theory comprises makes this approach to superstring field theory
the best one so far.

Green and Schwarz found a space-time supersymmetric action for the
superstring [10],
$$\eqalign{
S=\int d^2\sigma\{\,&\sqrt{-g}g^{\alpha\beta}\Pi^\mu_\alpha\Pi_{\mu\beta}
	\,-\,i\epsilon^{\alpha\beta}\partial_\alpha X^\mu
		(\overline\theta^1\gamma_\mu\partial_\beta\theta^1
		-\overline\theta^2\gamma_\mu\partial_\beta\theta^2)\cr
	&+\,i\epsilon^{\alpha\beta}
		\overline\theta^1\gamma^\mu\partial_\alpha\theta^1
		\overline\theta^2\gamma_\mu\partial_\beta\theta^2\,\}\cr}
$$
where
$$
\Pi^\mu_\alpha=\partial_\alpha X^\mu-
	i\overline\theta^A\gamma^\mu\partial_\alpha\theta^A
$$

This action has some interesting properties, that can as well be
analyzed in the simpler superparticle case, containing only
the first term in the action.
Its constraint structure is given by
$$\eqalign{
&L\equiv P^2\approx 0\cr
&\Phi\equiv p_\theta-iP_\mu\gamma^\mu\theta\approx 0\cr}
$$
There is one bosonic constraint, generating translations along the
world-line, and a fermionic spinor of local supersymmetry generators.
Due to $P^2=0$, the rank of
$$\{\Phi_a,\Phi_b\}=-2iP_\mu\gamma^\mu_{ab}$$
is 8 (out of 16).
The chiral spinor $\Phi$ thus contains eight \first\ class
constraints (``$\kappa$-symmetry'' [11]) and
eight \second\ class constraints.
There is no way (na\"\i vely) of eliminating covariantly the second
class constraints before quantization [12]. $\Phi$ is a chiral spinor,
transforming in the {\it irreducible} representation $16$ of $Spin(1,9)$,
and cannot be decomposed without giving up Lorentz covariance.
Exactly the same is true for the superstring.

The difficulty of getting rid of the second class constraints in a covariant
manner is closely connected to the problems with finding
covariant supersymmetric field theories in ten dimensions.

One can always choose a light-front gauge, where only the physical degrees
of freedom remain (no gauge invariance). Then the step to
field theory is straightforward [13].

A couple of approaches exist, where one tries to deal with the constraint
structure of the Green-Schwarz string by modifying it. Siegel
proposed [14]
that only the first class constraints should be kept, by only demanding
(in the superparticle version) $\Psi\equiv P_\mu(\gamma^\mu\Phi)\approx 0$, and
then introducing additional bosonic constraints removing fermionic
degrees of freedom. The problem here is that the spinor $\Psi$ has an
infinite level of reducibility. It is not clear how to treat the infinite
tower of ghosts that arise. Similar approaches are advocated in [15].

Much of the original work in supertwistors was motivated by the observation
that they have the potential of solving the problem with separation of
the fermionic constraints in \first\ and \second\ class parts.
Let me therefore briefly describe the fundamental ideas of division
algebra twistors
for massless bosonic particles and superparticles, and then discuss
application to string theory.

The classical dimensionalities of the superstring are $D=3,4,6,10$.
The gamma matrix identities
$(\lambda_1\gamma_\mu\lambda_2)\gamma^\mu\lambda_3+\hbox{cycl.}=0$
needed are directly related to the existence of
the (alternative) division algebras $\K_\nu=\R$, $\C$, $\H$ and $\O$.
More specifically:

\centerline{Existence of Clifford algebra $\leftrightarrow$ alternativity,}

\centerline{$v^\mu=\lambda\gamma^\mu\lambda$ lightlike
	$\leftrightarrow$ division property $|ab|=|a||b|$.}

This opens the way to twistor transformations of the lightlikeness
constraints in these dimensionalities:
$$P^\mu=\half\lambda\gamma^\mu\lambda\quad\Longleftrightarrow\quad
	P^{a\dot a}=\lambda^a\lambda^{\dagger\dot a}$$

The lightlike directions form the sphere $S^\nu$. The spinor, modulo
$\R_+$, lies on $S^{2\nu-1}$, where $\nu=D-2$. The spinor $\lambda$
is a two-component object
$\lambda=\left[\lambda^1\,\,\lambda^2\right]^t$
with entries in $\K_\nu$, transforming under
$SL(2;\K_\nu)\approx Spin(1,\nu+1)$.
A vector is a hermitean matrix.
$$
v=\left[\matrix{v^+&v^*\cr v&v^-}\right]$$

The twistor transform [16] from $\lambda$ to $P$ is the {\it Hopf map}
$S^{2\nu-1}\rightarrow S^\nu$
with fiber $S^{\nu-1}$.
The realization of the last (octonionic) Hopf map relies on the
understanding of $S^7$ as ``almost a Lie group'' [17].

The twistor transform can be extended to superparticles [18], and it
{\it solves the \second\ class constraint problem!}
The fermionic variables become a Lorentz scalar element of the division
algebra, except in D=10, where such things do not exist, and the fermion
carries a vector representation, which actually can be identified as
the fermionic varibles of the spinning particle [19].
All phase space variables sit in a representation of
$OSp(1|4;\K_\nu)\quad (\nu\neq 8)$, which is the superconformal
group [20] in D=3,4,6.

Strings are not (space-time) conformally invariant, except in the
zero tension limit. A twistor transformation of the lightlikeness
condition
$(\partial X)^2=0$ as $\partial X=\lambda\lambda^\dagger$
introduces \second\ class constraints between $\lambda$ and its
canonical momentum $\omega$, since $X$ already spans the entire
phase space for the left-(right-)moving sector.

Simple counting of the number of degrees of freedom in D=10 gives
$$8\,(\hbox{phys.})=2\!\times\!16-2\!\times\!1\,(\hbox{Vir.})-
	2\!\times\!7\,(\hbox{affine }S^7)
	-n\quad\Longrightarrow\quad n=8\,\,,$$

so there are must be 8 \second\ class constraints for the
bosonic twistor string
in D=10. These constraints are quite analogous to the fermionic
ones in the space-time picture of the superstring. The problem
with fermionic constraints can be solved, but it reappears in
the bosonic sector! This is quite general for twistor formulation of strings.
The problem is not universally recognized, but
seems to be generic in the sense that it appears as soon as
chiral spinors form part of phase space.
It is not at all clear whether the problem can be circumvented.
We are not in the position that we dare to formulate a no-go theorem.

Even though the problem I have pointed out seems to be a very severe one
concerning the prospect of finding a covariant quantization scheme for
superstrings,
there is a lot of very interesting structure in twistor superstrings.
Different versions exist, each with its own advantages.
\item{1.}{N=8 superconformal algebra (based on $S^7$) as a gauge
	group [21].
	This formulation is manifestly ``octonionic'', which I consider
	as fundamental.}
\item{2.}{N=8 superfield formulation where the $\kappa$
	symmetry is identified
	as a local world-sheet supersymmetry [22].
	The r\^ole of superconformal symmetry here is less clear.}

\noindent It is likely that there exists an N=8 superfield formulation of 1.,
but the theory of N=8 superconformal field theory is unexplored.

A very interesting and intriguing observation is that
the Green-Schwarz superstring gauge-fixed to the light-cone exhibits
an N=8 superconformal symmetry [23].
The spin 2 and spin 3/2 generators are remnants of the Virasoro and
local fermionic constraints, but the $S^7$ generators can not be traced
back to a symmetry in this way. It is tempting to think that it
is actually a sign of a gauge symmetry in an action where the \second\
class constraints reflect a partial choice of gauge.
This still remains a speculation -- we have not been able to find
such a formulation.
\vskip\baselineskip
\noindent{\twelvebf References.}\vskip.5\baselineskip

\item{ 1.} {S.~Hwang, \it Phys.Rev. \bf D28\rm (1983)2614.
		}
\item{ 2.} {W.~Siegel and B.~Zwiebach, \it Nucl.Phys.
	\bf B263\rm (1986)105,\brk
 	T.~Banks and M.F.~Peskin, in ``Unified String Theories'',
	World Scientific (1986),\brk
	A.~Neveu and P.~West, \it Nucl.Phys. \bf B268\rm (1986)125.\brk
	E.~Witten, \it Nucl.Phys. \bf B268\rm (1986)253
		}
\item{ 3.} {B.~Zwiebach, Lectures at Les Houches Summer School 1992,
	{\tt hep-th/9305026},\brk
	E.~Witten, \it Phys.Rev. \bf D46\rm (1992)5467,\brk
	A.~Sen and B.~Zwiebach, \it Nucl.Phys. \bf 414\rm (1994)649.
		}
\item{ 4.} {P.A.M.~Dirac, \it Rev.Mod.Phys. \bf 21\rm (1949)392.
		}
\item{ 5.} {I.A.~Batalin and E.S.~Fradkin,
		\it Nucl.Phys. \bf B279\rm (1987)514.
		}
\item{ 6.} {P.~Ramond, \it Phys.Rev. \bf D3\rm (1971)2415,\brk
	A.~Neveu and J.H.~Schwarz, \it Nucl.Phys. \bf B31\rm (1971)1109.
		}
\item{ 7.} {L.~Brink, P.~DiVecchia and P.~Howe,
		\it Phys.Lett. \bf 65B\rm (1976)471.
		}
\item{ 8.} {F.~Gliozzi, J.~Scherk and D.~Olive,
		\it Nucl.Phys. \bf B122\rm (1977)253.
		}
\item{ 9.} {D.~Friedan, E.~Martinec and S.~Shenker,
		\it Phys.Lett. \bf 160B\rm (1985)55,\brk
	E.~Witten, \it Nucl.Phys. \bf B276\rm (1986)291\brk
	C.R.~Preitschopf, C.B.~Thorn and S.A.~Yost,
		\it Nucl.Phys. \bf B337\rm (1990)363.
		}
\item{10.} {M.B.~Green and J.H.~Schwarz,
		\it Phys.Lett. \bf 136B\rm (1984)367.
		}
\item{11.} {W.~Siegel, \it Phys.Lett. \bf 128B\rm (1983)397.
		}
\item{12.} {I.~Bengtsson and M.~Cederwall, G\"oteborg-ITP-84-21.
		}
\item{13.} {M.B.~Green, J.H.~Schwarz and L.~Brink,
		\it Nucl.Phys. \bf B219\rm (1983)437.
		}
\item{14.} {W.~Siegel, \it Nucl.Phys. \bf B263\rm (1986)93.
		}
\item{15.} {R.~Kallosh, \it Phys.Lett. \bf 225B\rm (1989)49.
		}
\item{16.} {R.~Penrose and M.A.H.~McCallum,
		\it Phys.Rep. \bf 6\rm (1972)241,\brk
	I.~Bengtsson and M.~Cederwall, \it Nucl.Phys.
		\bf B302\rm (1988)81,\brk
	N.~Berkovits, \it Phys.Lett. \bf 247B\rm (1990)45,\brk
	M.~Cederwall, \it J.Math.Phys. \bf 33\rm (1992)388.
		}
\item{17.} {M.~Cederwall and C.R.~Preitshopf, {\tt hep-th/9309030},
		\it Commun.Math.Phys. \rm in press.
		}
\item{18.} {A.~Ferber, \it Nucl.Phys. \bf B132\rm (1978)55,\brk
	T.~Shirafuji, \it Progr.Theor.Phys. \bf 70\rm (1983)18,\brk
	I.~Bengtsson and M.~Cederwall, \it Nucl.Phys.
		\bf B302\rm (1988)81,\brk
	N.~Berkovits, \it Phys.Lett. \bf 247B\rm (1990)45,\brk
	M.~Cederwall, \it J.Math.Phys. \bf 33\rm (1992)388.
		}
\item{19.} {M.~Cederwall, \it Mod.Phys.Lett. \bf A9\rm (1994)967.
		}
\item{20.} {A.~Sudbery, \it J.Phys. \bf A17\rm (1984)939.
		}
\item{21.} {N.~Berkovits, \it Nucl.Phys. \bf B358\rm (1991)169,\brk
	M.~Cederwall and C.R.~Preitshopf, {\tt hep-th/9309030},
		\it Commun.Math.Phys. \rm in press.
		}
\item{22.} {D.P.~Sorokin, V.I.~Tkach and D.V.~Volkov,
		\it Mod.Phys.Lett. \bf A4\rm (1989)901,\brk
	M.~Tonin, \it Int.J.Mod.Phys. \bf A7\rm (1992)6013,\brk
	F.~Delduc, A.~Galperin, P.~Howe and E.~Sokatchev,
		\it Phys.Rev. \bf D47\rm (1992)578,\brk
	I.A.~Bandos, M.~Cederwall, D.P.~Sorokin and D.V.~Volkov,
		{\tt hep-th/9403181},\brk\indent
		\it Mod.Phys.Lett. \bf A \rm in press.
		}
\item{23.} {L.~Brink, M.~Cederwall and C.~R.~Preitschopf,
	\it Phys.Lett. \bf 311B\rm (1993)76.
		}
\vskip\baselineskip

\noindent The reference list, and also the text, gives a very fragmented
rendering of the contributions to a vast subject. The author sincerely
apologizes for that.

\end